\documentclass[twocolumn,showpacs,preprintnumbers,amsmath,amssymb]{revtex4}
\usepackage{docs}

\usepackage{graphicx}
\usepackage{dcolumn}

\usepackage{bm}

\begin{document}

\title{Stability of phantom wormholes}

\author{Francisco S. N. Lobo}
\email{flobo@cosmo.fis.fc.ul.pt}

\affiliation{Centro de Astronomia
e Astrof\'{\i}sica da Universidade de Lisboa, \\
Campo Grande, Ed. C8 1749-016 Lisboa, Portugal}


\begin{abstract}

It has recently been shown that traversable wormholes may be
supported by phantom energy. In this work phantom wormhole
geometries are modelled by matching an interior traversable
wormhole solution, governed by the equation of state $p=\omega
\rho$ with $\omega<-1$, to an exterior vacuum spacetime at a
finite junction interface. The stability analysis of these phantom
wormholes to linearized spherically symmetric perturbations about
static equilibrium solutions is carried out.
A master equation dictating the stability regions is deduced, and
by separating the cases of a positive and a negative surface
energy density, it is found that the respective stable equilibrium
configurations may be increased by strategically varying the
wormhole throat radius.
The first model considered, in the absence of a thin shell, is
that of an asymptotically flat phantom wormhole spacetime. The
second model constructed is that of an isotropic pressure phantom
wormhole, which is of particular interest, as the notion of
phantom energy is that of a spatially homogeneous cosmic fluid,
although it may be extended to inhomogeneous spherically symmetric
spacetimes.

\end{abstract}

\pacs{04.20.Gz, 04.20.Jb, 98.80.Es}

\maketitle


\section{Introduction}

It is now generally accepted that the Universe is undergoing an
accelerated phase of
expansion~\cite{Riess2,Perlmutter,Bennet,Hinshaw}, where the scale
factor obeys $\ddot{a}>0$. This cosmic acceleration is one of the
most challenging current problems in cosmology. Several
candidates, responsible for this expansion, have been proposed in
the literature, namely, dark energy models, generalizations of the
Chaplygin gas, modified gravity and scalar-tensor theories,
tachyon scalar fields and braneworld models, amongst others. The
dark energy models are parametrized by an equation of state given
by $\omega =p/\rho$, where $p$ is the spatially homogeneous
pressure and $\rho$ is the dark energy density. For the cosmic
expansion, a value of $\omega<-1/3$ is required, as dictated by
the Friedman equation $\ddot{a}/a=-4\pi (p+\rho/3)$.
A specific exotic form of dark energy, denoted phantom energy, has
also been proposed, possessing the peculiar property of
$\omega<-1$. This parameter range is not excluded by observation,
and possesses peculiar properties, such as the violation of the
null energy condition and an infinitely increasing energy density,
resulting in a Big Rip, at which point the Universe blows up in a
finite time~\cite{Weinberg}. However, recent fits to supernovae,
CMB and weak gravitational lensing data indicate that an evolving
equation of state $\omega$ crossing the phantom divide $-1$, is
mildly favored, and several models have been proposed in the
literature
\cite{Zhang2,Zhang3,Periv1,Wei-Cai,Li-Feng,Stef,Periv2,Feng,Vikman,Tsujikawa,Sami}.
In particular, models considering a redshift dependent equation of
state, $\omega(z)$, provide significantly ameliorated fits to the
most recent and reliable SN Ia supernovae Gold dataset
\cite{Riess3}.

As the phantom energy equation of state, $p=\omega \rho$ with
$\omega<-1$, violates the null energy condition, $p+\rho<0$, the
fundamental ingredient to sustain traversable
wormhole~\cite{Morris,mty,Visser}, one now has at hand a possible
source for these exotic spacetimes. In fact, this possibility has
recently been explored \cite{Sushkov,Lobo-phantom}, and it was
shown that traversable wormholes can be theoretically supported by
phantom energy. However, a subtlety needs to be pointed out, as
emphasized in Refs. \cite{Sushkov,Lobo-phantom}. The notion of
phantom energy is that of a homogeneously distributed fluid. When
extended to inhomogeneous spherically symmetric spacetimes, the
pressure appearing in the equation of state is now a radial
pressure, and the transverse pressure is then determined via the
field equations. In this context, it is interesting to note that
wormhole solutions with an isotropic pressure were found in
\cite{Lobo-phantom}, although these geometries are not
asymptotically flat. Sushkov, in Ref. \cite{Sushkov}, found
wormhole geometries by considering specific choices for the
distribution of the energy density, and in \cite{Lobo-phantom}, a
complementary approach was traced out, by imposing appropriate
choices for the form function and/or the redshift function, and
the stress-energy tensor components were consequently determined.
In Ref. \cite{Lobo-phantom} it was also shown, using the ``volume
integral quantifier'' \cite{VKD1,VKD2}, that these geometries can
be theoretically constructed with infinitesimal amounts of
averaged null energy condition violating phantom energy.

It is also of a fundamental importance to investigate the
stability of these phantom wormhole geometries (It is also
interesting to note that a stability analysis of a specific class
of traversable wormholes was carried out in Ref. \cite{Armen}, in
a rather different context). As in Ref. \cite{Lobo-phantom}, we
shall model these spacetimes by matching an interior traversable
wormhole geometry with an exterior Schwarzschild vacuum solution
at a junction interface \cite{Lobo-CQG,LLQ,Lobo,LL-PRD}. In this
work, we analyze the stability of these phantom wormholes to
linearized perturbations around static solutions. Work along these
lines was done by considering thin-shell Schwarzschild wormholes,
using the cut-and-paste technique \cite{Poisson}. It was later
shown that the inclusion of a charge \cite{Eiroa} and of a
cosmological constant \cite{LC-CQG} significantly increases the
stable equilibrium configurations found in Ref. \cite{Poisson}.
The advantage of this analysis resides in using a parametrization
of the stability of equilibrium, so that there is no need to
specify a surface equation of state. Note that the stability
analysis of these thin-shell wormholes to linearized spherically
symmetric perturbations about static equilibrium solutions was
carried out by assuming that the shells remain transparent under
perturbation \cite{Ishak}. This amounts to considering specific
spacetimes that do not contribute with the momentum flux term in
the conservation identity, which provides the conservation law for
the surface stress-energy tensor. The inclusion of this term,
corresponding to the discontinuity of the momentum impinging on
the shell, severely complicates the analysis. However, we shall
follow the approach of Ishak and Lake \cite{Ishak}, with the
respective inclusion of the momentum flux term, and deduce a
master equation responsible for dictating the stability
equilibrium configurations for the specific phantom wormhole
geometries found in Ref. \cite{Lobo-phantom}. We shall separate
the cases of a positive and a negative surface energy density, and
find that the stability may be significantly increased by varying
the wormhole throat.

This paper is outlined in the following manner. In Section II, we
present solutions of a phantom energy traversable wormhole. In
Section III, we outline a general linearized stability analysis
procedure, and deduce a master equation dictating stable
equilibrium configurations. We then apply this analysis to phantom
wormhole geometries and determine their respective stability
regions. Finally in Section IV, we conclude.

\section{Phantom energy traversable wormholes}

\subsection{Field equations}

The interior wormhole spacetime is given by the following metric
\cite{Morris}
\begin{equation}
ds^2=-e ^{2\Phi(r)}\,dt^2+\frac{dr^2}{1- b(r)/r}+r^2 \,(d\theta
^2+\sin ^2{\theta} \, d\phi ^2) \label{metricwormhole}  \,,
\end{equation}
where $\Phi(r)$ and $b(r)$ are arbitrary functions of the radial
coordinate, $r$, denoted as the redshift function and the form
function, respectively \cite{Morris}. The wormhole throat is
located at $b(r_0)=r=r_0$. For the wormhole to be traversable, one
must demand that there are no horizons present, which are
identified as the surfaces with $e^{2\Phi}\rightarrow 0$, so that
$\Phi(r)$ must be finite everywhere. The condition $1-b/r>0$ is
also imposed.
The stress-energy tensor components are given by (with $c=G=1$)
\begin{eqnarray}
\rho(r)&=&\frac{1}{8\pi} \,\frac{b'}{r^2} \label{rhoWH} \,, \\
p_r (r)&=&\frac{1}{8\pi} \left[-\frac{b}{r^3}+2
\left(1-\frac{b}{r} \right) \frac{\Phi'}{r} \right] \label{prWH}
\,,
\end{eqnarray}
\begin{eqnarray}
p_t(r)&=&\frac{1}{8\pi} \left(1-\frac{b}{r}\right)\Bigg[\Phi ''+
(\Phi')^2- \frac{b'r-b}{2r(r-b)}\Phi'
    \nonumber    \\
&&-\frac{b'r-b}{2r^2(r-b)}+\frac{\Phi'}{r} \Bigg] \label{ptWH}\,,
\end{eqnarray}
where $\rho(r)$ is the energy density; $p_r (r)$ the radial
pressure; and $p_t(r)$ the transverse pressure. The conservation
of the stress-energy tensor, $T^{\mu\nu}{}_{;\nu}=0$, provides us
with the following relationship
\begin{equation}
p_r'=\frac{2}{r}\,(p_t-p_r)-(\rho +p_r)\,\Phi '
\label{prderivative} \,.
\end{equation}

A fundamental ingredient of traversable wormholes and phantom
energy is the violation of the null energy condition (NEC), which
is defined as $T_{\mu\nu}k^{\mu}k^{\nu} \geq 0$, where $k^\mu$ is
any null vector and the $T_{\mu\nu}$ the stress-energy tensor.
Note that for phantom energy, governed by the equation of state
$\omega=p/\rho$ with $\omega<-1$, one readily verifies that the
NEC is violated, i.e., $p+\rho<0$. For wormhole spacetimes,
consider an orthonormal reference frame with $k^{\hat{\mu}}=(1,\pm
1,0,0)$, so that we have
\begin{equation}\label{NECthroat}
T_{\hat{\mu}\hat{\nu}}k^{\hat{\mu}}k^{\hat{\nu}}=
\frac{1}{8\pi}\,\left[\frac{b'r-b}{r^3}+
2\left(1-\frac{b}{r}\right) \frac{\Phi '}{r} \right]  \,.
\end{equation}
Thus, using the flaring out condition of the throat,
$(b-b'r)/2b^2>0$ \cite{Morris,Visser}, and considering the finite
character of $\Phi(r)$, we verify that evaluated at the throat the
NEC is violated, i.e.,
$T_{\hat{\mu}\hat{\nu}}k^{\hat{\mu}}k^{\hat{\nu}}<0$. Matter that
violates the NEC is denoted as {\it exotic matter}.

Note that the notion of phantom energy is that of a homogeneously
distributed cosmic fluid. However, as emphasized in
\cite{Sushkov,Lobo-phantom}, it may be extended to inhomogeneous
spherically symmetric spacetimes by regarding that the pressure in
the equation of state $p=\omega \rho$ is now a radial pressure
$p_r$. The transverse pressure $p_t$ may then be determined from
the field equation, in particular, from Eq. (\ref{ptWH}). Thus, to
find phantom energy traversable wormhole spacetimes, we use the
equation of state $p_r=\omega \rho$ with $\omega<-1$, representing
phantom energy, and thus deduce the following relationship
\begin{equation}
\Phi'(r)=\frac{b+\omega rb'}{2r^2\,\left(1-b/r \right)} \,,
            \label{EOScondition}
\end{equation}
by taking into account Eq. (\ref{rhoWH}) and Eq. (\ref{prWH}).

To model a traversable wormhole, one now considers appropriate
choices for $b(r)$ and/or $\Phi(r)$. Note that this is necessary
as we only have four equations, namely, Eqs.
(\ref{rhoWH})-(\ref{ptWH}), and Eq. (\ref{EOScondition}), with
five unknown functions of $r$, i.e., $\rho(r)$, $p_r(r)$,
$p_t(r)$, $b(r)$ and $\Phi(r)$. We shall only consider form
functions of the type $b'(r)>0$, as in cosmology the phantom
energy density is considered positive. Now, using the flaring out
condition evaluated at the throat \cite{Morris,Visser}, we also
have the condition $b'(r_0)<1$.

One may construct asymptotically flat spacetimes, in which
$b(r)/r\rightarrow 0$ and $\Phi\rightarrow 0$ as $r\rightarrow
\infty$. However, one may also consider solutions with a cut-off
of the stress-energy, by matching the interior solution to an
exterior vacuum spacetime, at a junction interface, $a$. For
simplicity, in this paper, we shall consider that the exterior
spacetime is the Schwarzschild solution, so that the matching
occurs at a junction interface, $r=a$, situated outside the event
horizon, i.e., $a>r_b=2M$, in order to avoid a black hole
solution.

\subsection{Specific phantom wormhole models}

The physical properties and characteristics of specific phantom
energy traversable wormhole models were analyzed in Ref.
\cite{Lobo-phantom}, by considering asymptotically flat spacetimes
and by imposing an isotropic pressure. Using the ``volume integral
quantifier'' it was found that it is theoretically possible to
construct these geometries with vanishing amounts of averaged null
energy condition violating phantom energy. Specific wormhole
dimensions and the traversal velocity and time were also deduced
from the traversability conditions for a particular wormhole
geometry. We shall briefly summarize two specific phantom wormhole
models, found in Ref. \cite{Lobo-phantom}, and for which we shall
further analyze the respective stable equilibrium configurations.

\subsubsection{Asymptotically flat spacetimes}

To construct an asymptotically flat wormhole solution
\cite{Lobo-phantom}, consider $\Phi(r)={\rm const}$. Thus, from
Eq. (\ref{EOScondition}) one obtains
\begin{equation}
b(r)=r_0(r/r_0)^{-1/\omega}  \,,
\end{equation}
so that $b(r)/r=(r_0/r)^{(1+\omega)/\omega}\;\rightarrow 0$ for
$r\rightarrow \infty$. We also verify that
$b'(r)=-(1/\omega)(r/r_0)^{-(1+\omega)/\omega}$, so that at the
throat the condition $b'(r_0)=1/|\omega|<1$ is satisfied.

The stress-energy tensor components are given by
\begin{eqnarray}
p_r(r)&=&\omega \rho(r) =-\frac{1}{8\pi
r_0^2}\left(\frac{r_0}{r}\right)^{3+\frac{1}{\omega}} \,,
    \\
p_t(r)&=&\frac{1}{16\pi r_0^2}\left(\frac{1+\omega}{\omega}\right)
\left(\frac{r_0}{r}\right)^{3+\frac{1}{\omega}} \,.
\end{eqnarray}
Thus, determining the parameter $\omega$ from observational
cosmology, assuming the existence of phantom energy, one may
theoretically construct traversable phantom wormholes by
considering the above-mentioned form function and a constant
redshift function.

\subsubsection{Isotropic pressure, $p_r=p_t=p$}

It was found that considering an isotropic pressure, $p_r=p_t=p$,
for $\Phi(r)$ to be finite one cannot construct asymptotically
flat traversable wormholes \cite{Lobo-phantom}. By taking into
account the form function given by $b(r)=r_0\,(r/r_0)^\alpha$,
with $0<\alpha<1$, and using Eq. (\ref{prderivative}) and Eq.
(\ref{rhoWH}), one finds that the redshift function is given by
\begin{equation}
\Phi(r)=\left(\frac{3\omega+1}{1+\omega}\right)\;\ln
\left(\frac{r}{r_0}\right) \,,
\end{equation}
where the relationship $\alpha=-1/\omega$ is imposed (see Ref.
\cite{Lobo-phantom} for details). The stress-energy tensor
components are provided by
\begin{eqnarray}
p(r)&=&\omega \rho(r) =-\frac{1}{8\pi
r_0^2}\left(\frac{r_0}{r}\right)^{3+\frac{1}{\omega}} \,.
\end{eqnarray}

As noted above, the spacetime is not asymptotically flat.
Nevertheless, one may match the interior wormhole solution to an
exterior vacuum spacetime at a finite junction surface.

\section{Stability analysis}

\subsection{Junction conditions}

We shall model specific phantom wormholes by matching an interior
traversal wormhole geometry, satisfying the equation of state
$p_r=\omega \rho$ with $\omega<-1$, with an exterior Schwarzschild
solution at a junction interface $\Sigma$, situated outside the
event horizon, $a>r_b=2M$.

Using the Darmois-Israel formalism \cite{Darmois,Israel}, the
surface stress-energy tensor, $S^i{}_j$, at the junction interface
$\Sigma$ is provided by the Lanczos equations
\begin{equation}
S^{i}_{\;j}=-\frac{1}{8\pi}\,(\kappa ^{i}_{\;j}-\delta
^{i}_{\;j}\kappa ^{k}_{\;k})  \,,
      \label{Lanczos}
\end{equation}
where $\kappa_{ij}$ is the discontinuity of the extrinsic
curvatures across the surface $\Sigma$, i.e.,
$\kappa_{ij}=K_{ij}^{+}-K_{ij}^{-}$. The extrinsic curvature is
defined as $K_{ij}=n_{\mu;\nu}\,e^{\mu}_{(i)}e^{\nu}_{(j)}$, where
$n^{\mu}$ is the unit normal $4-$vector to $\Sigma$, and
$e^{\mu}_{(i)}$ are the components of the holonomic basis vectors
tangent to $\Sigma$.

Taking into account the wormhole spacetime metric
(\ref{metricwormhole}) and the Schwarzschild solution, the
non-trivial components of the extrinsic curvature are given by
\begin{eqnarray}
K ^{\tau
\;+}_{\;\;\tau}&=&\frac{\frac{M}{a^2}+\ddot{a}}{\sqrt{1-\frac{2M}{a}+\dot{a}^2}}
\;,  \label{Kplustautau2}\\
K ^{\tau \;-}_{\;\;\tau}&=&\frac{\Phi'
\left(1-\frac{b}{a}+\dot{a}^2
\right)+\ddot{a}-\frac{\dot{a}^2(b-b'a)}{2a(a-b)}}{\sqrt{1-\frac{b(a)}{a}+\dot{a}^2}}
\;, \label{Kminustautau2}
\end{eqnarray}
and
\begin{eqnarray}
K ^{\theta
\;+}_{\;\;\theta}&=&\frac{1}{a}\sqrt{1-\frac{2M}{a}+\dot{a}^2}\;,
 \label{Kplustheta2}\\
K ^{\theta
\;-}_{\;\;\theta}&=&\frac{1}{a}\sqrt{1-\frac{b(a)}{a}+\dot{a}^2}
\;.  \label{Kminustheta2}
\end{eqnarray}

The Lanczos equation, Eq. (\ref{Lanczos}), then provide us with
the following expressions for the surface stresses
\begin{eqnarray}
\sigma&=&-\frac{1}{4\pi a} \left(\sqrt{1-\frac{2M}{a}+\dot{a}^2}-
\sqrt{1-\frac{b(a)}{a}+\dot{a}^2} \, \right)
    \label{surfenergy}   ,\\
{\cal P}&=&\frac{1}{8\pi a} \Bigg[\frac{1-\frac{M}{a}
+\dot{a}^2+a\ddot{a}}{\sqrt{1-\frac{2M}{a}+\dot{a}^2}}
   \nonumber    \\
&&-\frac{(1+a\Phi') \left(1-\frac{b}{a}+\dot{a}^2
\right)+a\ddot{a}-\frac{\dot{a}^2(b-b'a)}{2(a-b)}}{\sqrt{1-\frac{b(a)}{a}+\dot{a}^2}}
\, \Bigg]         \,,
    \label{surfpressure}
\end{eqnarray}
where $\sigma$ and ${\cal P}$ are the surface energy density and
the tangential surface pressure, respectively.

We shall make use of the conservation identity, which is obtained
from the second contracted Gauss-Kodazzi equation or the ``ADM"
constraint $G_{\mu \nu}e^{\mu}_{(i)}n^{\nu}=K^j_{i|j}-K,_{i}$ with
the Lanczos equations, and is given by
\begin{eqnarray}\label{conservation}
S^{i}_{j|i}=\left[T_{\mu \nu}e^{\mu}_{(j)}n^{\nu}\right]^+_-\,.
\end{eqnarray}
The momentum flux term in the right hand side corresponds to the
net discontinuity in the momentum which impinges on the shell.

Using $S^{i}_{\tau|i}=-\left[\dot{\sigma}+2\dot{a}(\sigma +{\cal
P} )/a \right]$, Eq. (\ref{conservation}) provides us with
\begin{equation}
\sigma'=-\frac{2}{a}\,(\sigma+{\cal P})+\Xi
  \,,\label{consequation2}
\end{equation}
where $\Xi$, defined for notational convenience, is given by
\begin{eqnarray}
\Xi=-\frac{1}{4\pi a^2} \left[\frac{b'a-b}{2a\left(1-\frac{b}{a}
\right)}+a\Phi' \right] \sqrt{1-\frac{b}{a}+\dot{a}^2} \,.
       \label{H(a)}
\end{eqnarray}

For self-completeness, we shall also include the $\sigma +{\cal
P}$ term, which is given by
\begin{eqnarray}
\sigma+{\cal P}&=& \frac{1}{8\pi a}\Bigg[\frac{(1-a\Phi')
\left(1-\frac{b}{a}+\dot{a}^2
\right)-a\ddot{a}+\frac{\dot{a}^2(b-b'a)}{2(a-b)}}
{\sqrt{1-\frac{b(a)}{a}+\dot{a}^2}}
    \nonumber  \\
&&-
 \frac{1-\frac{3M}{a}
+\dot{a}^2-a\ddot{a}}{\sqrt{1-\frac{2M}{a}+\dot{a}^2}} \, \Bigg]
    \,.
    \label{s+P}
\end{eqnarray}
Thus, taking into account Eq. (\ref{s+P}), and the definition of
$\Xi$, we verify that Eq. (\ref{consequation2}) finally takes the
form
\begin{eqnarray}
\sigma'&=&\frac{1}{4\pi a^2} \Bigg(\frac{1-\frac{3M}{a}
+\dot{a}^2-a\ddot{a}}{\sqrt{1-\frac{2M}{a}+\dot{a}^2}}
\nonumber   \\
&& - \frac{1-\frac{3b}{2a}+\frac{b'}{2}+\dot{a}^2
-a\ddot{a}}{\sqrt{1-\frac{b}{a}+\dot{a}^2}} \, \Bigg) \,,
     \label{sigma'WH}
\end{eqnarray}
which, evaluated at a static solution $a_0$, shall play a
fundamental role in determining the stability regions. Note that
Eq. (\ref{sigma'WH}) can also be deduced by taking the radial
derivative of the surface energy density, Eq. (\ref{surfenergy}).

\subsection{Equation of motion}

Rearranging Eq. (\ref{surfenergy}) into the form
\begin{equation}
\sqrt{1-\frac{2M}{a}+\dot{a}^2}=
\sqrt{1-\frac{b(a)}{a}+\dot{a}^2}\,- 4\pi \sigma a   \,,
\end{equation}
we deduce the thin shell's equation of motion, i.e.,
\begin{equation}
\dot{a}^2 + V(a)=0\,,
\end{equation}
with the potential given by
\begin{equation}\label{potential}
V(a)=1+\frac{2M\,b(a)}{m_s^2}
-\left[\frac{m_s}{2a}+\frac{\left(M+\frac{b(a)}{2} \right)}{m_s}
\right]^2   \,,
\end{equation}
where $m_s=4\pi \sigma a^2$ is the surface mass of the thin shell.
However, for computational purposes and notational convenience, we
define the following factors
\begin{eqnarray}
F(a)&=&1-\frac{b(a)/2+M}{a}
\,,      \label{factorF}   \\
G(a)&=&\frac{M-b(a)/2}{a} \,,
         \label{factorG}
\end{eqnarray}
so that the potential $V(a)$ takes the form
\begin{equation}
V(a)=F(a)-\left(\frac{m_s}{2a}\right)^2-\left(\frac{aG}{m_s}\right)^2
\,.
\end{equation}

Linearizing around a stable solution situated at $a_0$, we
consider a Taylor expansion of $V(a)$ around $a_0$ to second
order, given by
\begin{eqnarray}
V(a)&=&V(a_0)+V'(a_0)(a-a_0)
     \nonumber   \\
&&+\frac{1}{2}V''(a_0)(a-a_0)^2+O[(a-a_0)^3] \,.
\label{linear-potential}
\end{eqnarray}

The first and second derivatives of $V(a)$ are given by
\begin{eqnarray}
V'(a)&=&F'-2\left(\frac{m_s}{2a}\right)\left(\frac{m_s}{2a}\right)'
-2\left(\frac{aG}{m_s}\right)\left(\frac{aG}{m_s}\right)'
       \\
V''(a)&=&F''-2\left[\left(\frac{m_s}{2a}\right)'\right]^2-
2\left(\frac{m_s}{2a}\right)\left(\frac{m_s}{2a}\right)''
      \nonumber  \\
&&-2\left[\left(\frac{aG}{m_s}\right)'\right]^2
-2\left(\frac{aG}{m_s}\right)\left(\frac{aG}{m_s}\right)'' \,,
\end{eqnarray}
respectively. Evaluated at the static solution, at $a=a_0$, we
verify that $V(a_0)=0$ and $V'(a_0)=0$. From the condition
$V'(a_0)=0$, one extracts the following useful equilibrium
relationship
\begin{eqnarray}
\Gamma\equiv\left(\frac{m_s}{2a_0}\right)'
=\left(\frac{a_0}{m_s}\right)\left[
F'-2\left(\frac{a_0G}{m_s}\right)\left(\frac{a_0G}{m_s}\right)'\right]
  \,,
\end{eqnarray}
which will be used in determining the master equation, responsible
for dictating the stable equilibrium configurations.

The solution is stable if and only if $V(a)$ has a local minimum
at $a_0$ and $V''(a_0)>0$ is verified. The latter condition takes
the form
\begin{eqnarray}
\left(\frac{m_s}{2a}\right)\left(\frac{m_s}{2a}\right)''<\Psi
-\Gamma^2 \,,
     \label{masterequation}
\end{eqnarray}
where $\Psi$ is defined as
\begin{eqnarray}
\Psi=\frac{F''}{2}-\left[\left(\frac{aG}{m_s}\right)'\right]^2
-\left(\frac{aG}{m_s}\right)\left(\frac{aG}{m_s}\right)'' \,.
\end{eqnarray}

\subsection{The master equation}

Using $m_s=4\pi a^2 \sigma$, and taking into account the radial
derivative of $\sigma'$, Eq. (\ref{consequation2}) can be
rearranged to provide the following relationship
\begin{equation}
\left(\frac{m_s}{2a}\right)''= \Upsilon -4\pi \sigma'\eta \,,
     \label{cons-equation2}
\end{equation}
with the parameter $\eta$ defined as $\eta={\cal P}'/\sigma'$, and
$\Upsilon $ given by
\begin{equation}
\Upsilon \equiv \frac{4\pi}{a}\,(\sigma+{\cal P})+2\pi a \, \Xi '
\,.
\end{equation}
Equation (\ref{cons-equation2}) will play a fundamental role in
determining the stability regions of the respective solutions.
Note that the parameter $\sqrt{\eta}$ is normally interpreted as
the speed of sound, so that one would expect that $0<\eta \leq 1$,
based on the requirement that the speed of sound should not exceed
the speed of light. However, in the presence of exotic matter this
cannot naively de done so. Therefore, in this work the above range
will be relaxed. We refer the reader to Ref. \cite{Poisson} for an
extensive discussion on the respective physical interpretation of
$\eta$ in the presence of exotic matter.

We shall use $\eta$ as a parametrization of the stable
equilibrium, so that there is no need to specify a surface
equation of state. Thus, substituting Eq. (\ref{cons-equation2})
into Eq. (\ref{masterequation}), one deduces the master equation
given by
\begin{equation}
\sigma' \,m_s \,\eta_0 > \Theta\,,
\end{equation}
where $\eta_0=\eta(a_0)$ and $\Theta$ is defined as
\begin{equation}
\Theta \equiv \frac{a_0}{2\pi} \left(\Gamma^2-\Psi \right)
+\frac{1}{4\pi}\,m_s\,\Upsilon    \,.
       \label{master}
\end{equation}
Now, from the master equation we find that the stable equilibrium
regions are dictated by the following inequalities
\begin{eqnarray}
\eta_0 &>& \overline{\Theta}, \qquad {\rm if} \qquad \sigma'
\,m_s>0\,,      \label{stability1}
       \\
\eta_0 &<& \overline{\Theta}, \qquad {\rm if} \qquad \sigma'
\,m_s<0\,,       \label{stability2}
\end{eqnarray}
with the definition
\begin{eqnarray}
\overline{\Theta}\equiv \frac{\Theta}{\sigma' \,m_s}\,.
\end{eqnarray}

We shall now model the phantom wormhole geometries by choosing the
specific form and redshift functions considered in Ref.
\cite{Lobo-phantom}, and consequently determine the stability
regions dictated by the inequalities
(\ref{stability1})-(\ref{stability2}). In the specific cases that
follow, the explicit form of $\overline{\Theta}$ is extremely
messy, so that as in \cite{Ishak}, we find it more instructive to
show the stability regions graphically.

\subsection{Stability regions}

\subsubsection{Asymptotically flat spacetimes}

Consider the specific choices for the redshift and form functions
given by
\begin{eqnarray}
\Phi(r)&=&{\rm const}   \,,\label{redshift1}
     \\
b(r)&=&r_0(r/r_0)^{-1/\omega}  \,,
           \label{form1}
\end{eqnarray}
respectively. These are solutions to Eq. (\ref{EOScondition}), for
an asymptotically flat spacetime.

The factor related to the net discontinuity of the momentum flux
impinging on the shell, $\Xi$, is provided by
\begin{equation}
\Xi=\frac{1}{8\pi
a_0^2}\;\frac{\left(\frac{1+\omega}{\omega}\right)
\left(\frac{r_0}{a_0}\right)^{\frac{1+\omega}{\omega}}}
{\sqrt{1-\left(\frac{r_0}{a_0}\right)^{\frac{1+\omega}{\omega}}}}
 \,.
\end{equation}
The factor deduced from the equilibrium condition, $\Gamma$, is
given by
\begin{equation}
\Gamma=\frac{1}{2a_0}\;\left[\frac{1}{2}\frac{\left(\frac{1+\omega}{\omega}\right)
\left(\frac{r_0}{a_0}\right)^{\frac{1+\omega}{\omega}}}
{\sqrt{1-\left(\frac{r_0}{a_0}\right)^{\frac{1+\omega}{\omega}}}}
- \frac{\frac{M}{a_0}}{\sqrt{1-\frac{2M}{a_0}}} \right]
 \,.
\end{equation}

The radial derivative of the surface energy density, $\sigma'$,
evaluated at the static solution, which will be fundamental in
determining the stability regions, takes the following form
\begin{eqnarray}
\sigma'=\frac{1}{4\pi a_0^2}
\Bigg(\frac{1-\frac{3M}{a_0}}{\sqrt{1-\frac{2M}{a_0}}} -
\frac{1-\left(\frac{1+\omega}{2\omega}\right)
\left(\frac{r_0}{a_0}\right)^{\frac{1+\omega}{\omega}}}
{\sqrt{1-\left(\frac{r_0}{a_0}\right)^{\frac{1+\omega}{\omega}}}}
\, \Bigg) \,.
     \label{specific-sigma'}
\end{eqnarray}

We shall not write down the explicit forms of the remaining
functions, i.e., $\Upsilon$, $\Psi$ and $\Theta$, as they are
extremely lengthy. However, the stability regions shall be shown
graphically.

To determine the stability regions of this solution, we shall
separate the cases of $b(a_0)<2M$ and $b(a_0)>2M$. From Eq.
(\ref{surfenergy}) and the definition of $m_s=4\pi a_0^2 \sigma$,
this corresponds to $m_s >0$ and $m_s <0$, respectively. Here, we
shall relax the condition that the surface energy density be
positive, as in considering traversable wormhole geometries, one
is already dealing with exotic matter. Note that for $\sigma<0$,
the weak energy condition is readily violated.

For $b(a_0)<2M$, i.e., for a positive surface energy density, and
using the form function, Eq. (\ref{form1}), we need to impose the
condition $r_0<2M$, so that the junction radius lies outside the
event horizon, $a_0>2M$. Thus, the junction radius lies in the
following range
\begin{equation}
2M< a_0 <2M \left(\frac{2M}{r_0}\right)^{-(1+\omega)}  \,.
     \label{flat-range}
\end{equation}
For a fixed value of $\omega$, we verify that as $r_0 \rightarrow
0$, then $a_0 \rightarrow \infty$. The range decreases, i.e., $a_0
\rightarrow 2M$, as $r_0 \rightarrow 2M$. Note that by fixing
$r_0$ and decreasing $\omega$, the range of $a_0$ is also
significantly increased.
\begin{figure}[t]
\centering
  \includegraphics[width=2.4in]{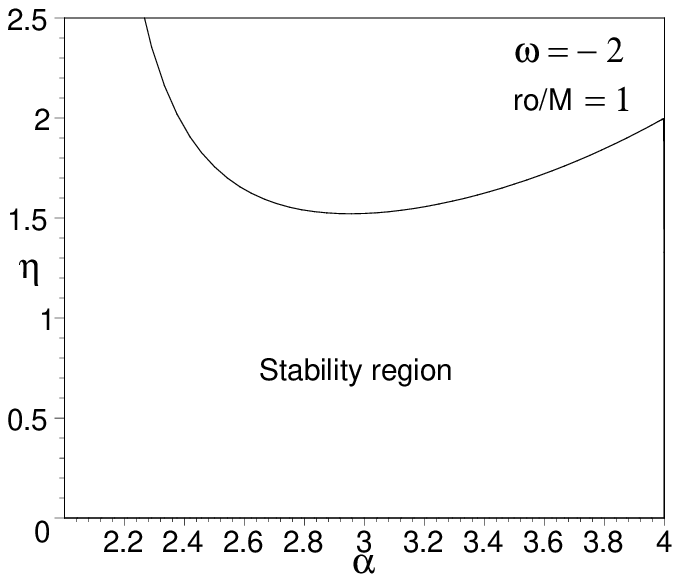}
  \hspace{0.4in}
  \includegraphics[width=2.4in]{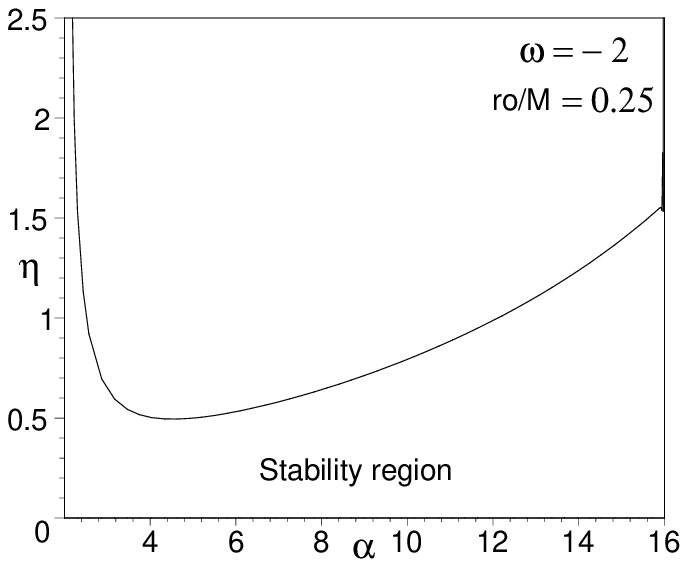}
  \caption{Plots for a positive surface energy density, i.e.,
  $b(a_0)<2M$. We have defined $\alpha=a_0/M$, and considered
  $\omega=-2$ for both cases. The first plot is given by $r_0/M=1$,
  and the second by $r_0/M=0.25$.
  The stability regions are given below the solid curve.
  See the text for details.}
  \label{phantomWH}
\end{figure}

For a fixed value of the parameter, for instance $\omega=-2$, we
shall consider the following cases: $r_0/M=1.0$, so that
$2<a_0/M<4$; and for $r_0/M=0.25$, we have $2<a_0/M<16$. The
respective stability regions are depicted in Fig. \ref{phantomWH}.
From Eq. (\ref{specific-sigma'}) we find that $\sigma'<0$, and as
we are considering a positive surface energy density, this implies
$m_s \sigma'<0$. Thus, the stability regions, dictated by the
inequality (\ref{stability2}), lie beneath the solid lines in the
plots of Fig. \ref{phantomWH}. Note that for decreasing values of
$r_0/M$, despite the fact that the range of $a_0$ increases, the
values of $\eta_0$ are further restricted. Thus, adopting a
conservative point of view, using positive surface energy
densities, we note that stable phantom wormhole geometries may be
found well within the bound of $0<\eta_0 \leq 1$, and the
stability regions increase for increasing values of $r_0/M$.
\begin{figure}[t]
\centering
  \includegraphics[width=2.4in]{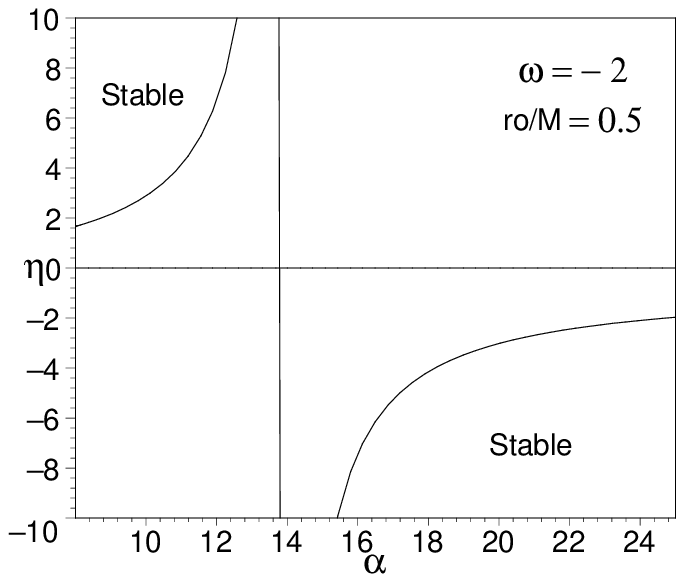}
  \hspace{0.4in}
  \includegraphics[width=2.4in]{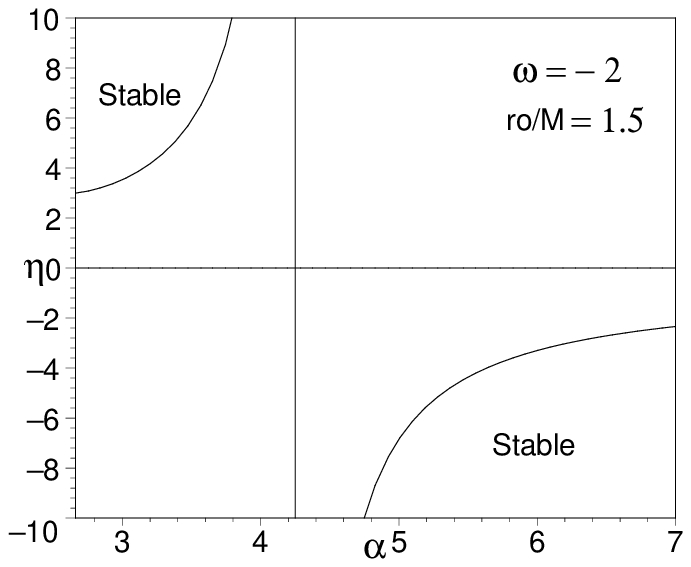}
  \caption{Plots for a negative surface energy density,
  considering $r_0/M<2$.
  We have defined $\alpha=a_0/M$, and considered
  $\omega=-2$ for both cases. The first plot is given by
  $r_0/M=0.5$, and the second by $r_0/M=1.5$.
  The stability regions are given above the first
  solid curve, and below the second solid curve.
  See the text for details.}
  \label{phantomWH2}
\end{figure}

For $b(a_0)>2M$, the surface mass of the thin shell is negative,
$m_s(a_0)<0$. We shall separate the cases of $r_0<2M$ and
$r_0>2M$.

If $r_0<2M$, the range of the junction radius is given by
\begin{equation}
a_0 >2M \left(\frac{2M}{r_0}\right)^{-(1+\omega)}  \,.
    \label{range2}
\end{equation}
For this specific case, $\sigma'$ possesses one real positive
root, $R$, in the range of Eq. (\ref{range2}), signalling the
presence of an asymptote, $\sigma'|_R=0$. We verify that
$\sigma'<0$ for $2M(2M/r_0)^{-(1+\omega)}<a_0 <R$, and $\sigma'>0$
for $a_0
>R$. Thus, the stability regions are given by
\begin{eqnarray}
\eta_0 &>& \overline{\Theta}, \quad {\rm if} \quad  2M
\left(\frac{2M}{r_0}\right)^{-(1+\omega)}<a_0 <R  \,,
         \label{flat-stability1}
       \\
\eta_0 &<& \overline{\Theta}, \quad {\rm if} \quad a_0 >R\,.
        \label{flat-stability2}
\end{eqnarray}

Consider for $\omega=-2$, the particular cases of $r_0/M=0.5$, so
that $a_0/M>8$, and $r_0/M=1.5$, so that $a_0/M>2.667$. The
asymptotes, $\sigma'|_R=0$, for these cases exist at $R/M \simeq
13.9$ and $R/M\simeq 4.24$, respectively. These cases are
represented in Fig. \ref{phantomWH2}. Note that for increasing
values of $r_0/M$, the range of $a_0$ decreases, and the values of
$\eta_0$ are less restricted. Thus, one may conclude that the
stability regions increase, for increasing values of $r_0/M$.

If $r_0>2M$, then obviously $a_0>r_0$. We verify that $\sigma'>0$,
and consequently $m_s\,\sigma'<0$, so that the stability region is
given by inequality (\ref{stability2}). We verify that the values
of $\eta_0$ are always negative. However, by increasing $r_0/M$,
the values of $\eta_0$ become less restricted, and the range of
$a_0$ decreases.

\subsubsection{Isotropic pressure, $p_r=p_t=p$}
\begin{figure}[t]
\centering
  \includegraphics[width=2.4in]{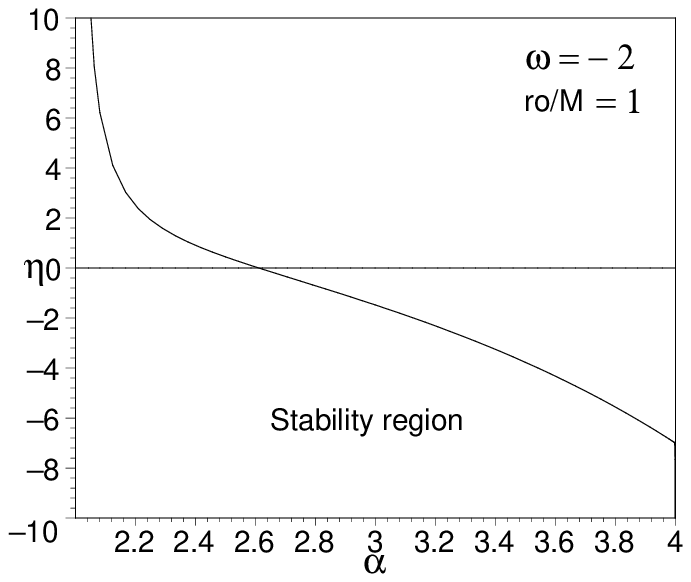}
  \hspace{0.4in}
  \includegraphics[width=2.4in]{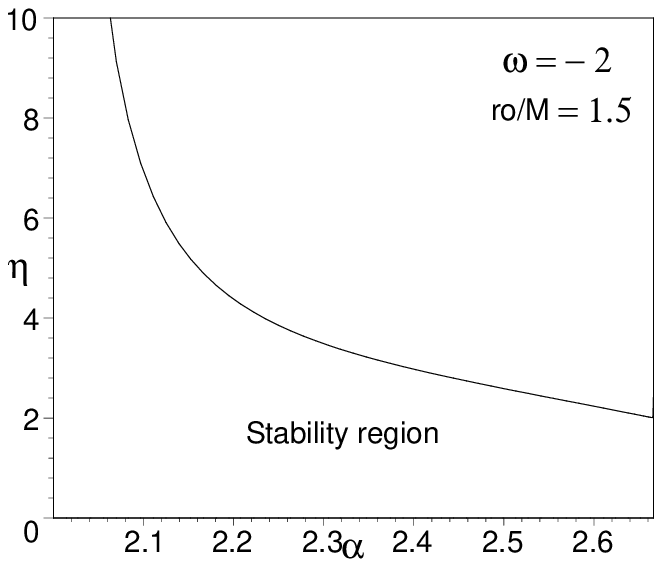}
  \caption{Plots for an isotropic pressure phantom wormhole.
  We have defined $\alpha=a_0/M$ and considered $\omega=-2$
  for both cases. For $b(a_0)<2M$, the condition $r_0<2M$ is imposed.
  The first plot is given by $r_0/M=1.0$, and the second by $r_0/M=1.5$.
  The stability regions are given below the solid curves.}
  \label{iso1}
\end{figure}

Consider the following functions
\begin{eqnarray}
\Phi(r)&=&\left(\frac{3\omega+1}{1+\omega}\right)\;\ln
\left(r/r_0\right)   \,,
    \\
b(r)&=&r_0\,(r/r_0)^{-1/\omega} \,,
\end{eqnarray}
which are solutions of a phantom wormhole possessing an isotropic
pressure \cite{Lobo-phantom}.

The factor related to the momentum flux term, $\Xi$, is given by
\begin{equation}
\Xi=\frac{1}{8\pi a_0^2}\;
\sqrt{1-\left(\frac{r_0}{a_0}\right)^{\frac{1+\omega}{\omega}}}
\left[
\frac{\left(\frac{1+\omega}{\omega}\right)
\left(\frac{r_0}{a_0}\right)^{\frac{1+\omega}{\omega}}}
{1-\left(\frac{r_0}{a_0}\right)^{\frac{1+\omega}{\omega}}}
-\frac{6\omega}{1+\omega} \right] \,.
\end{equation}
The $\Gamma$ and $\sigma'$ are identical to the previous case of
an asymptotically flat spacetime, and as before we shall not show
the specific forms of the remaining functions, as they are
extremely lengthly.
\begin{figure}[t]
\centering
  \includegraphics[width=2.4in]{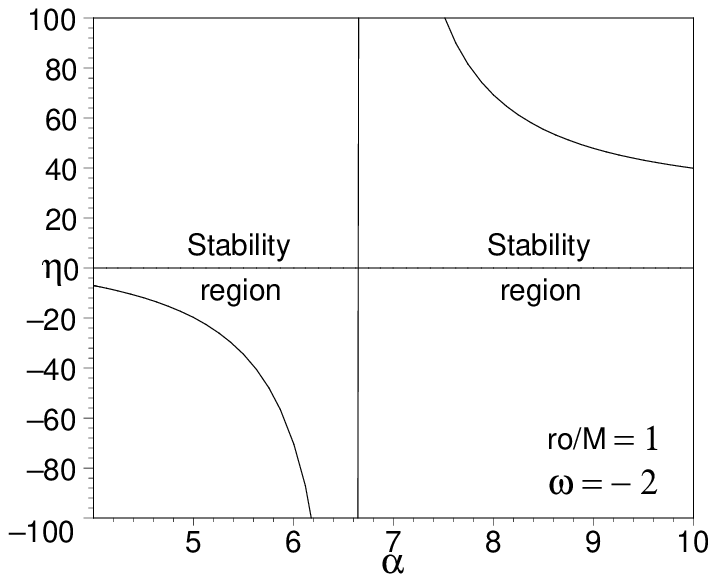}
  \hspace{0.4in}
  \includegraphics[width=2.4in]{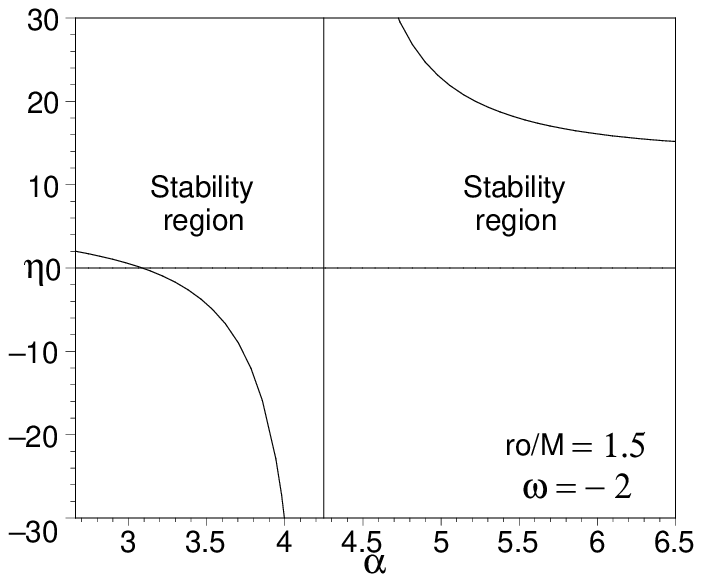}
  \caption{Plots for an isotropic pressure phantom wormhole,
  for $b(a_0)>2M$ and $r_0<2M$. We have defined $\alpha=a_0/M$ and
  considered $\omega=-2$ for both cases. The first plot is given
  by $r_0/M=1.0$, and the second by $r_0/M=1.5$.
  The stability regions are given above the first solid curve,
  and below the second solid curve.}
  \label{iso2}
\end{figure}

To determine the stability regions of this solution, as in the
previous case, we shall separate the cases of $b(a_0)<2M$ and
$b(a_0)>2M$.

For $b(a_0)<2M$, we have $m_s >0$, and the condition $r_0<2M$ is
imposed. Therefore, the junction radius lies in the same range as
the previous case, i.e., Eq. (\ref{flat-range}). We also verify
that $\sigma' <0$ in the respective range. Thus the stability
region is given by
\begin{equation}
\eta_0 < \overline{\Theta}, \quad {\rm if} \quad 2M< a_0 <2M
\left(\frac{2M}{r_0}\right)^{-(1+\omega)}  \,.
\end{equation}

Consider, for simplicity, $\omega=-2$, and the cases for $r_0/M=1$
and $r_0/M=1.5$ are analyzed in Fig. \ref{iso1}. The ranges are
given by $2<a_0/M<4$ and $2<a_0/M<2.667$, respectively. Note that
as $r_0/M$ decreases, the range of $a_0$ increases. However, the
values of the parameter $\eta_0$ become more restricted. Thus, one
may conclude that the stability regions increase, as $r_0/M$
increases.

For $b(a_0)>2M$, then $m_s(a_0)<0$. As before, we shall separate
the cases of $r_0<2M$ and $r_0>2M$. For $r_0<2M$, the range of
$a_0$ is given by $a_0 >2M (2M/r_0)^{-(1+\omega)}$, as in the
previous case of the asymptotically flat wormhole spacetime.

For this case $\sigma'$ also possesses one real positive root,
$R$, in the respective range. We have $\sigma'<0$ for
$(2M/r_0)^{-1/\omega}<a<R$, and $\sigma'>0$ for $a_0>R$. The
stability regions are also given by the conditions
(\ref{flat-stability1})-(\ref{flat-stability2}). We have
considered the specific cases of $r_0/M=1$ so that the respective
range is $a_0/M>4$; and $r_0/M=1.5$, so that $a_0/M>2.667$. The
asymptotes, $\sigma'|_R=0$, for these cases exist at $R/M\simeq
6.72$ and $R/M \simeq 4.24$, respectively. This analysis is
depicted in the plots of Fig. \ref{iso2}. Note that the plots
given by $\bar{\Theta}$ are inverted relatively to the
asymptotically flat spacetime. For decreasing values of $r_0/M$,
note that the value of the stability parameter $\eta_0$ becomes
less restricted and the range of the junction radius increases.
Thus, one may conclude that the stability regions increase for
decreasing values of $r_0/M$.

\begin{figure}[t]
\centering
  \includegraphics[width=2.4in]{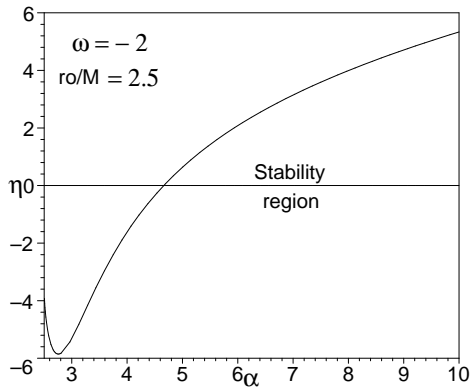}
  \caption{Plot for an isotropic pressure phantom wormhole,
  considering a negative surface energy density, with $r_0/M=2.5$.
  We have defined $\alpha=a_0/M$ and considered $\omega=-2$.
  The stability region is given below the solid curve.
  See the text for details.}
  \label{iso3}
\end{figure}

If $r_0>2M$, then $a_0>r_0$. We find that $\sigma'>0$, which
implies $m_s \sigma'<0$. Consider $\omega=-2$, and the specific
case of $r_0/M=2.5$, so that the stability region lies below the
solid line in Fig. \ref{iso3}. We also verify that for increasing
values of $r_0/M$, the values of the parameter $\eta_0$ become
further restricted. Thus, one may conclude that the stability
regions decrease for increasing values of $r_0/M$.

\section{Summary and Discussion}

As the Universe is probably constituted of approximately 70\% of
null energy condition violating phantom energy, this cosmic fluid
may be used as a possible source to theoretically construct
traversable wormholes. In fact, it was found that infinitesimal
amounts of phantom energy may support traversable wormholes
\cite{Lobo-phantom}. In this paper, we have modelled phantom
wormholes by matching an interior traversable wormhole geometry,
satisfying the equation of state $p=\omega \rho$ with $\omega<-1$,
to an exterior vacuum solution at a finite junction interface. We
have analyzed the stability of these phantom wormholes, an issue
of fundamental importance, to linearized perturbations around
static solutions, by including the momentum flux term in the
conservation identity. We have considered two particularly
interesting cases, namely, that of an asymptotically flat
spacetime, and that of an isotropic pressure wormhole geometry.
The latter solution is of particular interest, as the notion of
phantom energy is that of a spatially homogeneous cosmic fluid,
although it may be extended to inhomogeneous spherically symmetric
spacetimes. We have separated the cases of positive and negative
surface energy densities and found that the stable equilibrium
regions may be significantly increased by strategically varying
the wormhole throat. As we are considering exotic matter, we have
relaxed the condition $0 < \eta_0 \leq 1$, and found stability
regions for phantom wormholes well beyond this range. There are
several known examples of exotic $\eta_0 <0$ behavior, namely the
Casimir effect and the false vacuum \cite{Poisson}, so that one
cannot {\it a priori} impose $0 < \eta_0 \leq 1$ until a detailed
microphysical model of exotic matter is devised.

As emphasized in Ref. \cite{Lobo-phantom}, these stable phantom
wormholes have far-reaching physical and cosmological
implications. First, apart from being used for interstellar
travel, they may be transformed into
time-machines~\cite{mty,Visser}, consequently violating causality
with the associated time travel paradoxes. Relative to the
cosmological consequences, the existence of phantom energy
presents us with a natural scenario for traversal wormholes. It
was shown by Gonz\'{a}lez-D\'{i}az~\cite{gonzalez2}, that due to
the fact of the accelerated expansion of the Universe, macroscopic
wormholes could naturally be grown from the quantum foam. It was
shown that the wormhole's size increases by a factor which is
proportional to the scale factor of the Universe, and still
increases significantly if the cosmic expansion is driven by
phantom energy~\cite{diaz-phantom3}.
However, it was also found that using wormholes modelled by thin
shells accreting phantom energy \cite{Var-Israel}, the wormholes
become asymptotically comoving with the cosmological background as
the Big Rip is approached, so that the future of the universe is
shown to be causal.


\end{document}